\documentclass[twocolumn,secnumarabic,amssymb, nobibnotes, aps, prl, floatfix, superscriptaddress, nobalancelastpage]{revtex4-1}

\usepackage{xcolor}
\usepackage{fancyhdr}
\usepackage{lastpage}
\usepackage{amsmath}
\usepackage{graphicx}
\usepackage[utf8]{inputenc}
\usepackage{bm}
\usepackage{epsfig}

\usepackage{braket}
\usepackage{tensor}
\usepackage[version=4]{mhchem}
\usepackage{titlesec}
\usepackage{ifthen}
\usepackage{sidecap}
\usepackage{listings}
\usepackage[para,online,flushleft]{threeparttablex}

\usepackage{hyperref}
\usepackage[all]{hypcap}
\usepackage{float}

\usepackage{multibib}
\newcites{meth}{References}

\renewcommand{\figurename}{Figure}
\renewcommand{\tablename}{Table}

\setcitestyle{super}

\titleformat{\section}{\normalsize\raggedright\bfseries\fontsize{10}{12}}{\arabic{section}.}{1em}{}
\titleformat{\subsection}{\small\raggedright\bfseries}{\arabic{section}.}{1em}{}

\titlespacing\section{0pt}{12pt plus 4pt minus 2pt}{0pt plus 2pt minus 2pt}

\renewcommand{\vec}[1]{\mbox{\boldmath $#1$}}

\renewcommand{\vec}[1]{\mbox{\boldmath $#1$}}

\begin{document}

\begin{titlepage}
{\fontsize{26}{10} 
\textbf{\textcolor{black}{\flushleft 
Lightweight self-conjugate nucleus $^{80}$Zr}}}\\


{
A. Hamaker$^{1,2,3*}$, 
E. Leistenschneider$^{1,2,\dagger}$,
R. Jain$^{1,2,3}$,
G. Bollen$^{1,2,3}$,
S.A. Giuliani$^{1,4,5}$,
K. Lund$^{1,2}$,
W. Nazarewicz$^{1,3}$,
L. Neufcourt$^{1}$,
C. Nicoloff$^{1,2,3}$,
D. Puentes$^{1,2,3}$,
R. Ringle$^{1,2}$,
C.S. Sumithrarachchi$^{1,2}$,
I.T. Yandow$^{1,2,3}$
}

{
\fontsize{6}{10}{
\selectfont

$^{1}$Facility for Rare Isotope Beams, Michigan State University, East Lansing, Michigan 48824, USA.
$^{2}$National Superconducting Cyclotron Laboratory, Michigan State University, East Lansing, Michigan 48824, USA.
$^{3}$Department of Physics and Astronomy, Michigan State University, East Lansing, Michigan 48824, USA.
$^{4}$European Centre for Theoretical Studies in Nuclear Physics and Related Areas (ECT\textsuperscript{*}-FBK), Trento, Italy.
$^{5}$Department of Physics, Faculty of Engineering and Physical Sciences, University of Surrey, Guildford, Surrey GU2 7XH, United Kingdom.
$^*$Corresponding author: hamaker@nscl.msu.edu
$^{\dagger}$Current address: CERN, Geneva, Switzerland.
}
}

\end{titlepage}

{\bf Protons and neutrons in the atomic nucleus move in shells analogous to the electronic shell structures of atoms. Nuclear shell structure varies across the nuclear landscape due to changes of the nuclear mean field with the number of neutrons \vec{N} and protons \vec{Z}. These variations can be probed with mass differences. The \vec{N}=\vec{Z}=40 self-conjugate nucleus \textsuperscript{80}Zr is of particular interest as its proton and neutron shell structures are expected to be very similar, and its ground state  is highly deformed. In this work, we provide evidence for the existence of a deformed double shell closure in \textsuperscript{80}Zr through high precision Penning trap mass measurements of \textsuperscript{80-83}Zr. Our new mass values show that \textsuperscript{80}Zr is significantly lighter, and thus more bound than previously determined. This can be attributed to the deformed shell closure at \vec{N}=\vec{Z}=40 and the large Wigner energy. Our statistical Bayesian model mixing analysis employing several global nuclear mass models demonstrates difficulties with reproducing the observed  mass anomaly using current theory. }

Understanding the mechanisms of structural evolution, especially for exotic nuclei far from the beta stability line, is a major challenge in nuclear science\cite{Decadal2012,OTSUKA2020}. In this context, a rich territory for studies of basic nuclear concepts is the neutron-deficient region around mass number $A=80$\cite{ZirconiumRegion1988}. The nuclei in this region rapidly change their properties with proton and neutron numbers. Indeed, some of these nuclei are among the most deformed in the nuclear chart and exhibit  collective behaviour, while others show non-collective excitation patterns characteristic of spherical systems.

The appearance of strongly deformed configurations around \textsuperscript{80}Zr  has been attributed to the population of the intruder $g_{9/2}$ orbitals separated by the spherical $N=Z=40$ subshell closure from the upper-$pf$ shell. This particular shell structure results in coexisting configurations of different shapes predicted by theory\cite{Hamilton1984,NAZAREWICZ1985,PETROVICI1996,Gaudefroy2009,RODRIGUEZ2011,Kaneko2021}.
In particular, for the nucleus \textsuperscript{80}Zr, spherical and deformed (prolate, oblate, and triaxial) structures are expected to coexist at low energies, and their competition strongly depends on the size of the calculated spherical $N=Z=40$ gap\cite{Reinhard1999}. Experimentally, \textsuperscript{80}Zr
has a very large prolate quadrupole deformation  $\beta_2 \approx 0.4$\cite{Lister1987,Llewellyn2020}.
Within the mean-field theory, this  has been attributed to the appearance of the  large deformed gap at $N=Z=40$ in the deformed single-particle spectrum\cite{NAZAREWICZ1985}. Consequently, the nucleus  $^{80}$Zr can be viewed as a deformed doubly-magic  system. 

In addition to shape-coexistence effects, \textsuperscript{80}Zr is a great laboratory for isospin physics. Having equal number of protons and neutrons, this nucleus is self-conjugate; hence, it  offers a unique venue to study proton-neutron pairing, isospin breaking effects, and the Wigner energy reflecting an additional binding in self-conjugate nuclei and their neighbours\cite{SATULA1997,Bentley2013}.

The mass of an isotope is a sensitive indicator of the underlying shell structure as it reflects the net energy content of a nucleus, including the binding energy. Hence, doubly-magic nuclei are significantly lighter, or more bound, compared to their neighbours. Due to a lack of precision mass measurement data on \textsuperscript{80}Zr and its neighbours, it is difficult to characterize the size of the shell effect responsible for the large deformation of $^{80}$Zr. To this end, we performed high precision Penning trap mass spectrometry of four neutron-deficient zirconium isotopes -- \textsuperscript{80-83}Zr -- and analysed the local trends of the binding-energy surface by studying several binding-energy indicators. To quantify our findings, experimental patterns have been interpreted using global nuclear mass models augmented by a Bayesian model averaging analysis\cite{Neufcourt2020-limits}.

\section*{\textcolor{blue}{Experimental Procedure}}

\begin{figure*}[!htb]
        \includegraphics[width=1.0\linewidth]{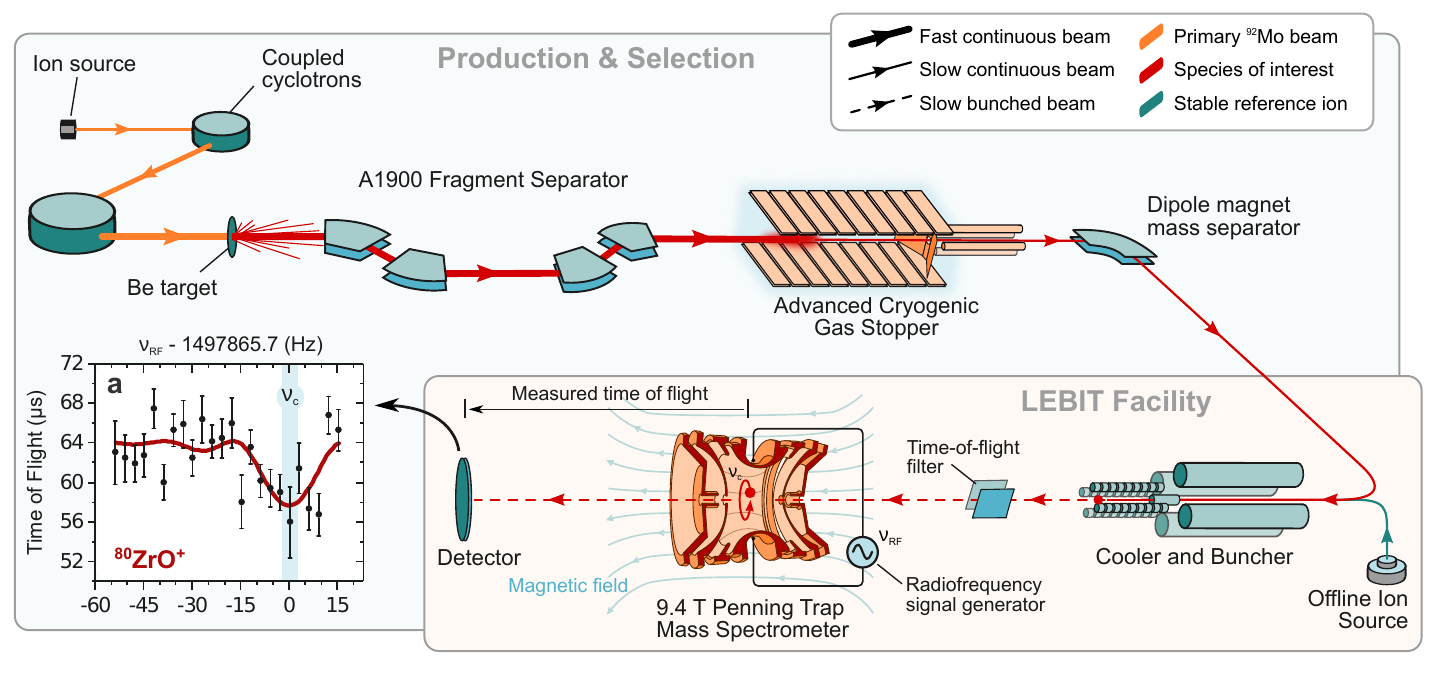}
        \caption{\textbf{The experimental procedure.} The relevant components of the experimental setup are displayed.  Panel \textbf{a} shows a sample time-of-flight spectrum of a \textsuperscript{80}Zr\textsuperscript{16}O\,\textsuperscript{+} molecular ion. The red curve is an analytical fit to the data\cite{Konig1995}. The error bars represent the statistical uncertainty of the 
        time-of-flight measurement, and the light blue band shows the $1\sigma$ uncertainty of the cyclotron frequency determination. See the main text and Methods for details.    }
        \label{fig:facility}
\end{figure*}
\textsuperscript{80-83}Zr are highly neutron-deficient unstable radioisotopes of zirconium with half-lives ranging between 4.6\,s and 42\,s\cite{ENSDF}, so they must be produced in specialized facilities and probed using fast and sensitive instrumentation. A schematic of the experimental setup and procedure is shown in Figure \ref{fig:facility}. The Zr isotopes were produced at the National Superconducting Cyclotron Laboratory's Coupled Cyclotron Facility via projectile fragmentation of a 140 MeV/u \textsuperscript{92}Mo primary beam that was impinged on a thin Be target. The produced Zr nuclei were separated from other fragments by the A1900 Fragment Separator\cite{MORRISSEY2003} and sent to the Advanced Cryogenic Gas Stopper\cite{LUND2020}, where they were stopped as ions. The ions were extracted from the gas stopper as a low energy (30 keV/Q) continuous beam and selected by their mass-to-charge ratio $(A/Q)$ using a dipole magnet. The ions were then sent to the Low Energy Beam and Ion Trap (LEBIT) facility\cite{RINGLE2013}. \textsuperscript{80,82}Zr ions were sent as singly charged oxides ($A/Q = 96, 98$ respectively); \textsuperscript{81,83}Zr ions were sent bare and doubly charged ($A/Q = 40.5, 41.5$ respectively).

Upon entering the LEBIT facility, the ions first passed through the cooler and buncher\cite{SCHWARZ2016} where they were accumulated, cooled, and released as short bunches to the LEBIT 9.4\,T Penning trap\cite{RINGLE2009}. A series of purification techniques described in Methods were employed to ensure nearly pure samples of the ion of interest were used for the measurement.  A schematic of the LEBIT setup is shown in Fig.~\ref{fig:facility}.

In the Penning trap, the mass $m_\text{ion}$ of an ion with charge $q$ was determined by measuring the cyclotron frequency $\nu_c = q\,B/ (2 \pi \, m_\text{ion})$ of the ion's motion about the trap's magnetic field, which has a strength $B$. The cyclotron frequency $\nu_c$ was measured using the time-of-flight ion cyclotron resonance (TOF-ICR) technique\cite{Konig1995}, illustrated in Fig.~\ref{fig:facility} and described in Methods. The theoretical line shapes\cite{Konig1995} for the TOF-ICR spectra were fit to the data allowing for the determination of the cyclotron frequency. A sample of a \textsuperscript{80}Zr\textsuperscript{16}O\,\textsuperscript{+} TOF-ICR spectrum and its theoretical line shape is shown in Fig.~\ref{fig:facility}a. 

Before and after each measurement of the ion of interest, measurements of a reference ion were performed to calibrate the magnetic field. The reference ions (\textsuperscript{41}K\textsuperscript{+}, \textsuperscript{85,87}Rb\textsuperscript{+}) were provided by an offline ion source. The masses of the ions of interest were obtained from the ratio ($R$) between cyclotron frequencies of the reference ion ($\nu_{c,\text{ref}}$) and the ion of interest: 
\begin{equation}
\label{eqn:freqratio} 
R  = \frac{\nu_{c,\text{ref}}}{\nu_{c}}  = \frac{m_\text{ion}/q_\text{ion}}{(m_\text{ref}-  q_\text{ref} \cdot  m_e)/q_{ref}} \, , \qquad \end{equation}
where $q_\text{ion}$ is the charge state of the ion of interest, and $m_e$ is the mass of the electron, while $m_\text{ref}$ and $q_\text{ref}$ are the atomic mass and charge state of the reference species. The atomic mass $m$ of the Zr isotope of interest is calculated from the mass of the measured ion, accounting for removed electrons and molecular counterparts, where applicable. The results of the measurements are displayed in Table~\ref{tab:results-thisexps} and compared to the Atomic Mass Evaluation of 2020 (AME20)\cite{AME20}. Further details on the measurement, calibration, and uncertainty determination procedures are given in Methods.

\begin{table*}[!htb]
\centering
  \caption{\textbf{Results from our mass measurements.} The mass excesses are relative to the atomic mass of the isotopes of interest. The average frequency ratio ($\bar{R}$) between the ion of interest (Ion)  and the reference ion (Ion Ref.) is presented. The results are also compared to the mass excesses recommended by the AME20\cite{AME20}. All mass excesses are in keV/c\textsuperscript{2}. $1\sigma$ uncertainties are shown in parenthesis.}
    {\begin{tabular}{c c c c c c c}
        Isotope &  Ion  & Ion Ref. & $\bar{R}$ & Mass Excess  & AME20\cite{AME20} & Difference \\    \hline
  $^{80}$Zr &  	$^{80}$Zr\,$^{16}$O\,$^{+}$		 &  $^{85}$Rb\,$^{+}$	 	&  1.129\,829\,01 (99)	& -55\,128 (80) &  -55\,517 (1\,500)$^\text{a}$ & 389 (1\,500)  \\

  $^{81}$Zr &  	$^{81}$Zr\,$^{2+}$	 &  $^{41}$K\,$^{+}$	&        0.987\,971\,08 (13)  & -57\,556 (10) & -57\,524 (92)	& -32 (93)  \\
 
  $^{82}$Zr &  	$^{82}$Zr\,$^{16}$O\,$^{+}$		 &  $^{87}$Rb\,$^{+}$	&  1.126\,770\,338 (31)   & -63\,618.6 (2.5)	& -63\,614.1 (1.6) & -4.5 (3.0) \\
 
  $^{83}$Zr &  	$^{83}$Zr\,$^{2+}$		 &  $^{41}$K\,$^{+}$	&  1.012\,274\,829\,7  (85)   & -65\,916.33 (65)	& -65\,911.7  (6.4) & -4.7 (6.5) \\ \hline
    \end{tabular}}%
   \\
    \footnotesize{ $^\text{a}$ Experimental result based on one $^{80}$Zr event\cite{Issmer1998}, not included in the AME20.}
  \label{tab:results-thisexps}%
\end{table*}

Our mass measurement results are in good agreement with the mass values recommended by AME20\cite{AME20}, and provide an improvement of one order of magnitude or more to the precision of the \textsuperscript{80,81,83}Zr masses. The AME20 values for \textsuperscript{81-83}Zr are derived mainly from previous high precision mass measurements. Penning trap mass measurements of \textsuperscript{82,83}Zr form the basis of the AME20 mass values for these isotopes\cite{KANKAINEN2006,Vilen2019}, while a recent storage ring measurement\cite{XING2018} dominates the AME20 mass of \textsuperscript{81}Zr. Our measurement of \textsuperscript{82}Zr has the largest discrepancy from AME20 with a value $1.5 \, \sigma$ lower. 
The mass of \textsuperscript{80}Zr listed in AME20 is an extrapolated value
calculated from neighbouring known nuclei using smooth
trends of the mass surface. It is worth noting that two previous mass measurements of \textsuperscript{80}Zr have not been included in the AME. A measurement with only a single event\cite{Issmer1998} yielded a mass uncertainty of 1.5\,MeV/$c^2$. The second measurement\cite{Lalleman2001}, albeit significantly more precise, has not been included in the AME because other isotopes measured in the same experiment were in disagreement with more recent high-precision results.


\section*{\textcolor{blue}{The Anomalous Mass of $^{80}$Zr}}

Our mass measurement of \textsuperscript{80}Zr reveals that this nucleus is significantly more bound than expected from systematic trends. Indeed, high-quality extrapolations of the mass surface towards $^{80}$Zr have been produced by the AME collaboration and others; this has been  especially motivated by the astrophysical significance of this nucleus for X-ray bursts\cite{Schatz2017}. 
Our mass value is 370 keV/$c^2$ more bound than the extrapolated value from AME20\cite{AME20}, and 950 keV/$c^2$ more bound than the Lanzhou extrapolated value\cite{XING2018}. 

To study the impact of our new masses we employed various binding-energy differences described in Methods. Along the $N=Z$ line, nuclei are known to be exceptionally well bound as neutrons and protons occupy the same shell model orbitals. Therefore, a useful indicator is the double mass difference $\delta V_{pn}$\cite{ZHANG1989,Stoitsov2007} defined in Methods.

\begin{figure}[!htb]
        \includegraphics[width=1.0\linewidth]{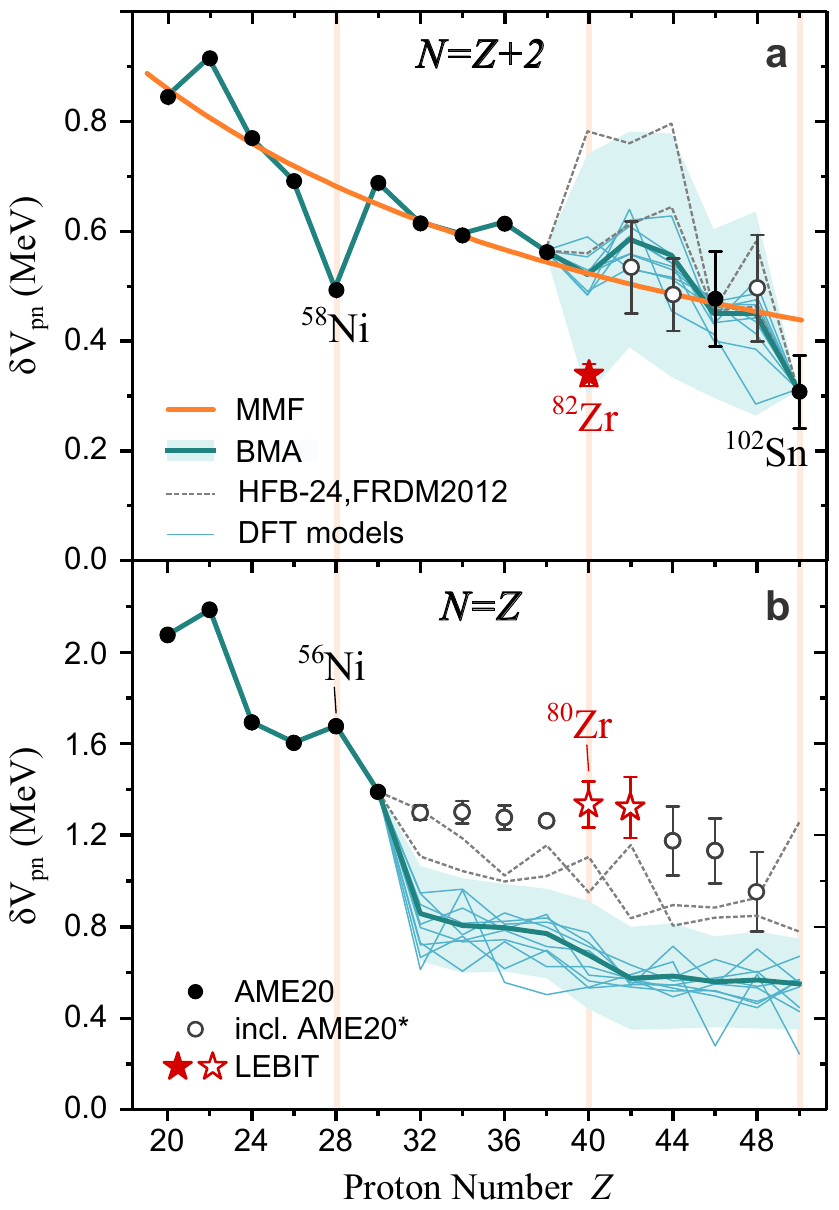}
        \caption{\textbf{Comparison of experimental results with theoretical predictions.} The effect of the anomalous mass of $^{80}$Zr on the mass indicator $\delta V_{pn}$:  a significant decrease from the baseline in the $N=Z+2$ sequence \textbf{(a)}, and a corresponding rise in the $N=Z$ sequence \textbf{(b)}, which mirrors the behaviour of other doubly-magic nuclei (e.g. $^{56}$Ni and $^{100}$Sn).  Black circles represent mass data from the AME20\cite{AME20}. Red stars include data from this work. Open symbols include mass extrapolations (AME20$^*$) from AME20\cite{AME20}. The MMF prediction is marked by an orange line in \textbf{(a)}.  
         The thick teal line is the BMA result based on several nuclear models (thin solid lines: DFT models; thin dashed lines: HFB-24 and FRDM2012 models that include the Wigner-energy correction), and the light teal band represents the uncertainty of the BMA approach. See Methods for details on the BMA.     }
        \label{fig:graphs}
\end{figure}

In Figure~\ref{fig:graphs}a and~\ref{fig:graphs}b, we show $\delta V_{pn}$ for  the $N=Z+2$ and $N=Z$ sequences, respectively. For nuclei away from  $N=Z$, the overall behaviour of $\delta V_{pn}$ is well described by the macroscopic mass formula\cite{Reinhard2006,Stoitsov2007} (MMF):   $\delta V_{pn} \approx 2(a_\text{sym} + a_\text{ssym}A^{-1/3})/A$, where $a_\text{sym}$ and $a_\text{ssym}$ are, respectively, symmetry and surface-symmetry energy coefficients. In the MMF plotted in Figure~\ref{fig:graphs}a, we employed $a_\text{sym} = 35$\,MeV and $a_\text{ssym} = -59$\,MeV, which were determined through a fit to the data neglecting the outliers at $A = 58, 82, 102$.
Along the $N=Z$ sequence, $\delta V_{pn}$ is strongly impacted by the Wigner energy\cite{SATULA1997}, whose behaviour is more convoluted. Moreover, mass data beyond  $N=Z$  are scarce in the investigated region.
Consequently, if some masses required for the $\delta V_{pn}$ determination were not experimentally available, we used the recommended values from AME20\cite{AME20} instead.

Although  $\delta V_{pn}$  is expected to vary smoothly overall,  fluctuations around the average trend carry important structural information\cite{ZHANG1989,Stoitsov2007,Bender2011}. Binding-energy outliers, especially those found in magic nuclei along the $N=Z$ line, result in $\delta V_{pn}$ deviations for both $N=Z$ and $N=Z+2$ sequences.  
Considering the $N=Z+2$ results with our new masses, the value of $\delta V_{pn}$ for $^{82}$Zr (which is reliant on the mass of $^{80}$Zr) is a clear outlier, being 185 keV lower than the MMF trend. This anomaly is similar to those  found in $^{58}$Ni and $^{102}$Sn, associated with the increased binding energies of the doubly-magic self-conjugate nuclei $^{56}$Ni and $^{100}$Sn. The increased binding energy of $^{80}$Zr also impacts the $N=Z$ trends resulting in increasing values of $\delta V_{pn}$ for Zr and Mo.

Analogous outliers can also be found inspecting other mass filters at $^{80}$Zr, such as the 2-proton shell gap $\delta_{2p}$, commonly employed in tests of shell closures\cite{Bender2002,Lunney2003}.  See Methods Figure 1 for additional discussion.

\begin{figure}[!htb]
        \includegraphics[width=1.0\linewidth]{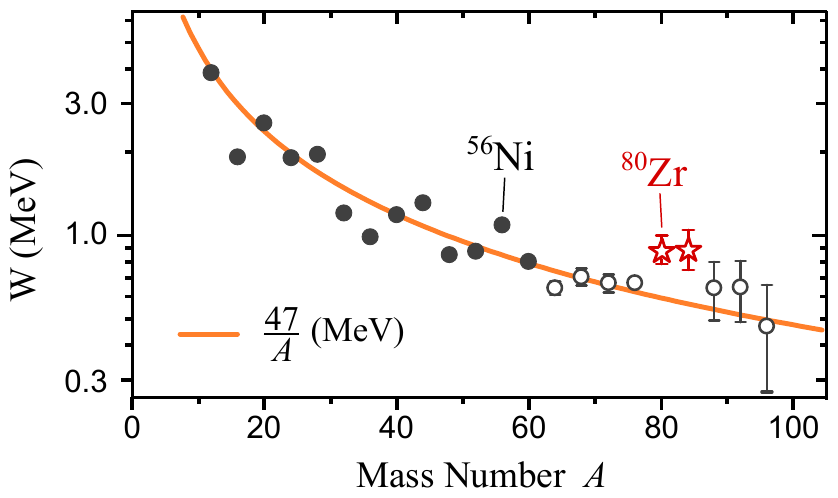}
        \caption{\textbf{Wigner energy.} 
        The Wigner-energy coefficient $W(A)$ extracted from $\delta V_{pn}$ values according to 
        Ref.\cite{SATULA1997}. Black circles represent the mass data from the AME20\cite{AME20}. Red stars include data from this work. Open symbols (AME20$^*$) include mass extrapolations. The average trend of Ref.\cite{SATULA1997} is shown by a thick line. }
        \label{fig:Wigner}
\end{figure}

The  results  shown in  Figure~\ref{fig:graphs} provide compelling empirical evidence for the existence of a deformed shell closure in $^{80}$Zr. One needs to bear in mind, however, that $^{80}$Zr is a self-conjugate system and some additional contribution to  its binding energy comes from the Wigner energy. Usually, the Wigner term in even-even nuclei is parametrized as $E_W=a_W|N-Z|/A$. As discussed in Ref.\cite{SATULA1997} and Methods, the Wigner-energy coefficient $W(A)=a_W/A$ can be empirically extracted from  the values of $\delta V_{pn}$.
Our data, shown in Figure~\ref{fig:Wigner}, indicates that the value of $W(A)$ at $^{80}$Zr and $^{56}$Ni is locally enhanced, contrary to the gradually decreasing trend for heavier $N = Z$ nuclei that is well captured by the value of $a_W=47$\,MeV obtained in Ref.\cite{SATULA1997}. A note of caution is in order: some contribution to the local increase of the empirical value of $W$ in $^{80}$Zr and $^{56}$Ni can be attributed to the enhanced binding due to their shell structure. The strength of the enhancement can be evaluated through another mass filter.

\begin{figure}[!htb]
        \includegraphics[width=1.0\linewidth]{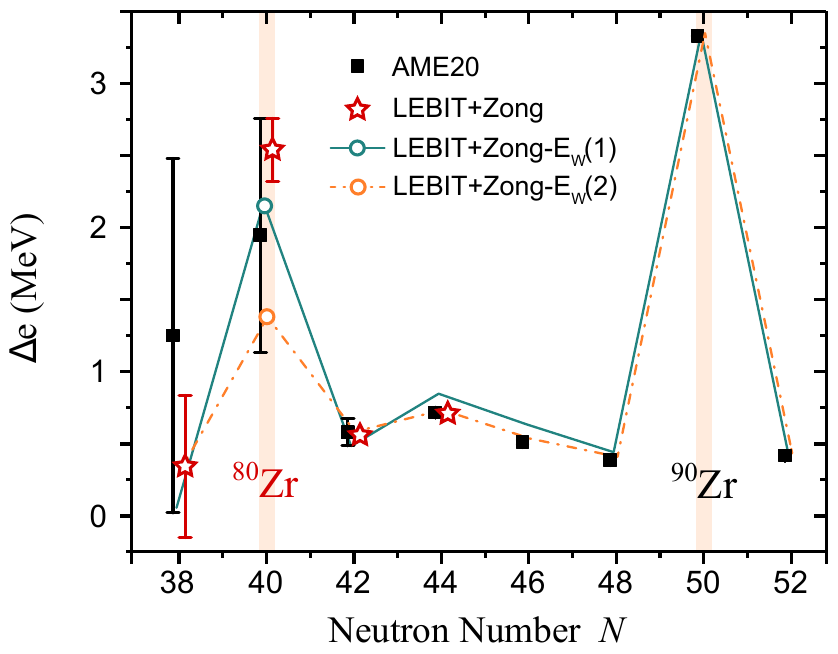}
        \caption{\textbf{Single-particle energy splitting.} 
        The empirical  single-particle energy gap $\Delta e(N)$ at the Fermi level for the chain of even-even  Zr isotopes extracted from nuclear binding energies according to
        Ref.\cite{Satula1998}. Black squares represent the mass data from the AME20\cite{AME20}. Open stars represent the data from this work augmented by mass extrapolations from Ref.\cite{Zong2020}. 
        These values of $\Delta e$ were further corrected by removing contributions from the Wigner energy term $E_W(1)$ (solid line; Ref.\cite{Goriely2013}) or $E_W(2)$ (dash-dotted line; Ref.\cite{SATULA1997}), cf. Methods for definition of $E_W$. Magic shell closures are seen at $N = 40$  and $N = 50$.
        }
        \label{fig:dE}
\end{figure}

Experimental masses offer a way to assess the size of the deformed $N=40$ single particle gap. This can be done  by employing the filter   $ \Delta e (N=2n)$\cite{Satula1998}, which provides an estimate of the  single-particle energy gap  $e_{n+1}-e_n$ at the Fermi level. 
Figure~\ref{fig:dE} shows $\Delta e$ for the Zr isotopic chain (cf.  Ref.\cite{Koszorus2021} for the applications of $\Delta e$ to the K and Ca chains). Some masses of proton-rich Zr isotopes needed to determine $\Delta e$ are not known experimentally; those have been taken from mass relations of mirror nuclei by Zong et al.\cite{Zong2020}. 
It is seen that $\Delta e$ reaches a maximum for $^{90}$Zr at the spherical magic number $N=50$ and a local maximum for $^{80}$Zr at the deformed magic number $N=40$.
Since the latter value can be affected by the Wigner energy, we removed the binding-energy contribution from $E_W$ by applying two models: $E_W(1)$\cite{Goriely2013} and $E_W(2)$\cite{SATULA1997}. The resulting correction to $\Delta e$ practically affects the $N=40$ value  only. As discussed in Methods, the expression $E_W(1)$ is well localized at $N=Z$ and reduces $\Delta e$ by about 300\,keV. The expression $E_W(2)$ decreases linearly with the neutron excess and the corresponding reduction of $\Delta e$ is about 1.1\,MeV.
Even in this case, the energy gap at $N=40$ is a factor of 2-3 larger than $\Delta e$ for $42\le N\le 48$. While the size of this gap is reduced as compared to the spherical $N=50$ gap, it is characteristic of a deformed shell closure. The strong shell effect comes from the self-conjugate nature of $^{80}$Zr as the deformed proton and neutron shell effects reinforce one another.

\section*{\textcolor{blue}{Bayesian Analysis of Mass Models}}

To obtain improved theoretical mass predictions in the \textsuperscript{80}Zr region,
we conducted a Bayesian statistical analysis 
combining Gaussian process extrapolation 
and Bayesian Model Averaging (BMA)\cite{BAND} 
of eleven theoretical global mass models following the same procedure as in Ref.~\cite{Neufcourt2020-limits}. The BMA framework uses the collective wisdom of the models, constrained by data, to make predictions and quantify uncertainties. Details on the individual models and the BMA methodology can be found in Methods.

The BMA predictions for $\delta V_{pn}$ are shown in Figure~\ref{fig:graphs}a,b. The predictions for $N=Z+2$ are well constrained outside the region $38<Z<50$ due to the wealth of experimental mass data. In the region $38 \leq Z \leq 50$, the BMA results  are consistent with the AME20 data and the MMF trend. At $Z=40$, the experimental $\delta V_{pn}$ value, which includes our new \textsuperscript{80,82}Zr mass results, falls just within the error band. The BMA result for $\delta V_{pn}$ along $N=Z$ line in the region $Z > 30$ does not agree with either the AME20 extrapolations or the new experimental value at $Z=40$. Two of the models, FRDM2012\cite{MOLLER2016} and HFB-24\cite{Goriely2013}, that include the phenomenological Wigner term perform slightly better than the DFT models. However, they still fall short of the experimental trends, most likely due to underestimated Wigner energy. Indeed, the value of $a_W$ in FRDM2012\cite{MOLLER2016} is 30\,MeV, which is significantly less than $a_W=47$\,MeV representing the average trend seen in Figure~\ref{fig:Wigner}. The Wigner energy  $E_W(1)$ of HFB-24 is even smaller.

\section*{\textcolor{blue}{Conclusions}}

The high precision mass measurements of \textsuperscript{80-83}Zr performed at the LEBIT facility allow a more detailed investigation of the mass surface in the region of strongly deformed nuclei with $N \approx Z \approx 40$. Our measurement reveals a significant enhancement in the binding energy of \textsuperscript{80}Zr. By considering binding-energy indicators, we attribute this enhancement to a deformed double shell closure and an increase in the Wigner energy of this exotic self-conjugate system. A Bayesian average based on eleven global mass models was unable to account for the new mass value of \textsuperscript{80}Zr as the Wigner-energy enhancement has not been taken into account microscopically. The comparisons to theory demonstrate the importance of accounting for the competition between deformation effects, isospin breaking effects, and proton-neutron pairing.

The interplay between theory and experiment was crucial in understanding this region of the nuclear chart. While the deformed shell gap at $N=Z=40$ was predicted over 30 years ago\cite{Hamilton1984,NAZAREWICZ1985}, a lack of precise experimental data prevented a quantitative assessment of the gap's size until now. To further refine the deformed shell closure, high precision mass measurements in this region are needed, which will be made possible with next-generation radioactive ion beam facilities and mass measurement techniques.

\bibliographystyle{naturemag}
\bibliography{library}

\titleformat{\subsection}[runin]{\small\raggedright\bfseries}{\arabic{section}.}{1em}{}

\clearpage
\section{\textcolor{blue}{METHODS}}

\setcounter{figure}{0}
\setcounter{table}{0}
\renewcommand{\figurename}{METHODS Figure}
\renewcommand{\tablename}{METHODS Table}
\renewcommand{\thefigure}{\arabic{figure}}
\renewcommand{\thetable}{\Roman{table}}
\renewcommand{\theHfigure}{METHODS Figure \thefigure}
\renewcommand{\theHtable}{METHODS Table \thetable}

\setcounter{equation}{0}

\textbf{\textcolor{blue}{The TOF-ICR technique for cyclotron frequency determination}}

In a Penning trap, an ion is confined in space by the superposition of a weak axially-harmonic electric potential and a strong homogeneous magnetic field, oriented in the axial direction. In the absence of the electric field, the ion performs a circular motion about the axis of the magnetic field at the cyclotron frequency $\nu_c$, whose measurement allows for the determination of the mass of the particle. The introduction of the electric field disturbs the cyclotron motion, which is split into two independent radial components: the reduced cyclotron and the much slower magnetron precession (with frequencies $\nu_+$ and $\nu_-$, respectively). The ``free" cyclotron frequency is determined by the measurement of the $\nu_c =  \nu_+ + \nu_-$ sideband. This quantity is nearly invariant with respect to fluctuations in the trapping electric field, which grants Penning trap mass spectrometry great accuracy\citemeth{Gabrielse2009}. 

In the TOF-ICR technique, the sideband is determined by applying an external quadrupole radiofrequency field (with frequency $\nu_{RF}$) to the ion that converts one eigenmotion into the other. The ion is initially prepared in a pure magnetron motion, which, at LEBIT, is done through the Lorentz steering technique\citemeth{Ringle2007_LS}. Upon the application of the external field, if the resonant condition $\nu_{RF} = \nu_+ + \nu_-$ is met, the conversion from pure magnetron motion to pure reduced cyclotron motion occurs. The conversion is probed by measuring the ion's time of flight from the trap to a microchannel plate detector outside of the magnetic field. If the ion in the trap is in a pure reduced cyclotron motion, which holds greater kinetic energy, the time of flight is significantly shorter. The LEBIT Facility section of Figure~\ref{fig:facility} provides a schematic of the TOF-ICR setup.

In a typical TOF-ICR procedure, $\nu_{RF}$ is scanned to characterize the resonant reduction of the time of flight, generating spectra such as the one shown in Figure~\ref{fig:facility}a. The width of the resonance, which determines the precision of the $\nu_c$ measurement, is inversely proportional to the time which the external excitation field is applied. In the measurements described herein, both continuous\cite{Konig1995} and Ramsey\citemeth{George2007} radiofrequency quadrupolar excitation schemes were used with excitation times ranging from 50\,ms to 1\,s. The cyclotron frequency is determined through an analytical fit to the time of flight spectrum, whose line shapes are described in the literature for both excitation schemes employed\cite{Konig1995}\textsuperscript{,}\citemeth{George2007}.

\textbf{\textcolor{blue}{Mass determination from cyclotron frequencies}}

Here we describe in greater detail the procedure employed to extract atomic mass values for the isotopes of interest from the measured cyclotron frequencies. As explained in the main text, each measurement of the ion of interest's cyclotron frequency $\nu_c$ was interleaved by measurements of the cyclotron frequency of the reference ion, $\nu_{c,\text{ref}}$. Reference ions were chosen as singly ionized species of widely available stable alkali atoms whose masses ($m_\text{ref}$) are well known in the literature\cite{AME20}, as well as whose $A/Q$ was close to the ion of interest to avoid large mass-dependent systematic shifts in the calibration procedure.    The frequency ratio (\ref{eqn:freqratio}) for each measurement of $\nu_c$ was calculated using the time-interpolated cyclotron frequency from the reference measurements to the time of the measurement of the ion of interest. In total, 3 measurements of $R$ were performed of the \textsuperscript{83}Zr\textsuperscript{2+}--\,\textsuperscript{41}K\textsuperscript{+} pair, 6 of the   \textsuperscript{82}ZrO\textsuperscript{+}--\,\textsuperscript{87}Rb\textsuperscript{+} pair, 5 of the \textsuperscript{81}Zr\textsuperscript{2+}--\,\textsuperscript{41}K\textsuperscript{+} pair, and 4 of the \textsuperscript{80}ZrO\textsuperscript{+}--\,\textsuperscript{85}Rb\textsuperscript{+} pair. The masses of each ion of interest ($m_\text{ion}$) were calculated through 
Eq.(\ref{eqn:freqratio}) using the average of multiple frequency ratios ($\bar{R}$), presented in Table \ref{tab:results-thisexps}. 

The atomic masses ($m$) of the Zr isotopes of interest were calculated using $m = m_\text{ion} + q \cdot m_e - m_\text{mol}$, where $m_\text{mol}$ is the atomic mass of the molecular counterpart ($^{16}$O in the case of $^{80,82}$Zr only).  The electron binding energies and molecular binding energies of \textsuperscript{80,82}ZrO\textsuperscript{+} were disregarded as they are on the order of eV, which is several orders of magnitude lower than the statistical uncertainty of the measurement. Mass excesses, defined as the difference between the atomic mass and the isotope's mass number, are reported in Table\,\ref{tab:results-thisexps} for the measured Zr isotopes.

\textbf{\textcolor{blue}{Evaluation of uncertainties}}

Uncertainties related to the extraction of cyclotron frequencies from the fits dominate the statistical error budget. Most systematic uncertainties in the measured frequency ratio scale linearly with the mass difference between the ion of interest and the reference ion. These systematic effects include magnetic field inhomogeneities, trap misalignment with the magnetic field, and anharmonic imperfections in the trapping potential\citemeth{BOLLEN1990}. The mass-dependent shifts in $\bar{R}$ have been measured at the LEBIT facility and found to be $\Delta \bar{R} = 2 \times 10^{-10}$/u \citemeth{GULYUZ2015}. This shift has been folded into the ratios and uncertainties reported in Table~\ref{tab:results-thisexps}.

Remaining systematic effects include nonlinear time-dependent changes in the magnetic field, relativistic effects on the cyclotron frequency, and ion-ion interactions in the trap. Previous work has shown that the effect of nonlinear magnetic field fluctuations on the individual ratios $R$ are less than $1 \times 10^{-9}$ over an hour \citemeth{RINGLE2007-3738Ca}. Measurement times ranged from three hours for \textsuperscript{80}Zr to fifteen minutes for \textsuperscript{83}Zr. This uncertainty was also folded into the ratio uncertainties though it had a negligible effect on the final error estimate. The relativistic effects on the cyclotron frequency ratio\citemeth{Brown1986} were negligible compared to the statistical uncertainty. Ion-ion interactions were minimized using several methods. Before entering the trap, the ion bunches from the cooler and buncher were purified using a time-of-flight filter to only allow ions with a specific mass-to-charge ratio to enter the trap. Once captured in the trap, ions were further purified against isobaric contamination using targeted dipole cleaning\citemeth{BLAUM2004} and the stored waveform inverse Fourier Transform (SWIFT) technique\citemeth{KWIATKOWSKI2015}. Additional ion-ion interactions were taken into account by performing a count-rate class analysis on each data set whenever possible\citemeth{BOLLEN1992}. The count-rate class analysis only led to a shift in the \textsuperscript{83}Zr ratio ($\Delta R = 9.8 (7) \times 10^{-9}$). This shift has been included in the value reported in Table~\ref{tab:results-thisexps}. Finally, Birge ratios were calculated to determine whether inner or outer uncertainties were reported for the final mass uncertainties\citemeth{BIRGE1932}.

\textbf{\textcolor{blue}{Binding-energy indicators}}

To extract quantities of interest for the experimental mass surface, we employ various  binding-energy differences (mass filters)\citemeth{JANECKE1985,JENSEN1984}. Those include:
\begin{description}
 \item[The double mass difference] $\delta V_{pn}$\cite{ZHANG1989,Stoitsov2007,Bender2011}:
\begin{equation}\label{eq:delVpn}
\begin{split}
    \delta V_{pn}(N,Z) &= \frac{1}{4}[B(N,Z)-B(N-2,Z) \\
    -  &B(N,Z-2)+B(N-2,Z-2)];
\end{split}
\end{equation}
\item[The Wigner energy coefficient] in an even-even nucleus with $N=Z=A/2$\cite{SATULA1997}:
\begin{equation}\label{eq:Wigner}
\begin{split}
   W(A)  &= \delta V_{pn}(A/2,A/2)\\
 - \frac{1}{2}&\left[\delta V_{pn}(A/2,A/2-2)+\delta V_{pn}(A/2+2,A/2)\right];
\end{split}
\end{equation}
\item[The two-proton shell gap] $\delta_{2p}$\cite{Bender2002,Lunney2003}:
\begin{equation}\label{eq:Dpp}
\delta_{2p}(N,Z)  = 2B(N,Z)-B(N,Z+2)-B(N,Z-2);
\end{equation}
\item[The three-point mass difference] $\Delta_n^{(3)}$\cite{Satula1998}:
\begin{equation}\label{eq:Dn}
\begin{split}
    \Delta_n^{(3)}(N,Z)=&\frac{(-1)^N}{2}[2B(N,Z) \\
    &-B(N-1,Z)-B(N+1,Z)];
\end{split}    
\end{equation}
\item[The single-particle energy splitting] $\Delta e$\cite{Satula1998}:
\begin{equation}\label{eq:DE}
\begin{split}
    \Delta  e(N,Z)=& e_{n+1}-e_n=2[\Delta_n^{(3)}(N=2n,Z) \\ -& \Delta_n^{(3)}(N=2n+1,Z)].
\end{split}    
\end{equation}
\end{description}
In the above equations,  $B$ is the (positive)  nuclear binding energy, obtained from the atomic mass of the nucleus.

\textbf{\textcolor{blue}{Wigner-energy parametrizations}} 

The Wigner energy contribution to the total binding energy
produces an additional binding for nuclei close to $N=Z$. In the HFB-24 mass model\cite{Goriely2013}, the Wigner term has been parametrized as:
\begin{equation}\label{EW1}
E_W(1)=V_We^{-\lambda_W\left(\frac{N-Z}{A}\right)^2} +V'_W |N-Z|e^{-\left(\frac{A}{A_0} \right)^2  },  
\end{equation}
where $V_W=1.8$\,MeV, $\lambda_W=380$, $V_W'=-0.84$\,MeV. and $A_0=26$. In  this model, $E_W$ rapidly decreases with $|N-Z|$ when moving away from the $N=Z$ line.  In the traditional parametrization of $E_W$, 
\begin{equation}\label{EW2}
E_W(2)=-a_W\frac{|N-Z|}{A}, 
\end{equation}
one assumes that $E_W=0$ at $N=Z$ and linearly decreases with the neutron excess. In this work, we adopt the value of $a_W=47$\,MeV from Ref.\cite{SATULA1997}.

\textbf{\textcolor{blue}{Nuclear models}}

In this study we considered 9 models based on nuclear density functional
theory (DFT): SkM\textsuperscript{*}\citemeth{BARTEL1982},
SkP\citemeth{DOBACZEWSKI1984}, SLy4\citemeth{CHABANET1995},
SV-min\citemeth{KLUPFEL2009}, UNEDF0\citemeth{KORTELAINEN2010},
UNEDF1\citemeth{KORTELAINEN2012}, UNEDF2\citemeth{KORTELAINEN2014},
D1M\citemeth{GORIELY2009}, and BCPM\citemeth{BALDO2013}. Two additional mass
models commonly used in nuclear astrophysics studies were also considered:
FRDM2012\cite{MOLLER2016} and HFB-24\cite{Goriely2013}.

Three of these models (SkM\textsuperscript{*}, UNEDF0 and
FRDM2012) predict large prolate ground-state deformation for $^{80}$Zr around $\beta_2=0.39$, in agreement with experiment.  HFB-24 predicts an oblate deformed ground state while all the remaining models predict a spherical ground state. Such variations in the predicted ground-state deformation are manifestations of near-lying coexisting configurations with  different shapes expected theoretically, as discussed in the main text. 
It is important to notice that while the relative position between the
different minima strongly depends on the underlying interaction\cite{Reinhard1999} and beyond-DFT correlations\cite{RODRIGUEZ2011}, the energy shifts between the deformed ground-state configuration and the spherical minimum is relatively small\cite{Reinhard1999}. As a consequence, the absolute impact of shape coexistence in the predicted mass value is expected to be minor and can be absorbed by the statistical correction.

\textbf{\textcolor{blue}{Bayesian Model Averaging}}

The  binding energies $B(N,Z)$ predicted by nuclear mass models were used to compute the two-proton separation energies:
\begin{align}
\begin{aligned}
		S_{2p} (N, Z) &= B(N, Z) - B(N, Z-2) \,,
		\label{eqs:separation_en}
  \end{aligned}
 \end{align} 
which were then used to compute  $\delta V_{pn}$  mass differences.

For each model employed, we construct the statistical emulator $\delta^{\rm em}_{S_{2p}}$ of separation energy residuals: 
\begin{equation}
	\delta^{\rm em}_{S_{2p}} (N,Z) := S^{exp}_{2p} (N,Z) - S^{th}_{2p}(N,Z) \,.
\end{equation}
The predicted separation energies are then given by
$S_{2p}^{em}(N,Z) = S_{2p}^{th}(N,Z) + \delta^{\rm em}_{S_{2p}}$. 
The training datasets were built from experimental masses from AME20 for even-even nuclei with $20 \leq Z \leq 50$ and the theoretical  mass tables. Seven nuclei
($^{48}$Ni, $^{54}$Zn, $^{84}$Zr, $^{86}$Mo, $^{90}$Ru, $^{92}$Ru, and
$^{94}$Pd) placed at the dataset outer boundary were excluded from the training set and used as independent testing data to compute the BMA evidence weights. Our dataset consists therefore of 152 points $(x_i, y_i)$, where $x:=(N_i, Z_i)$ and $y_i:=\delta^{\rm em}_{S_{2p}} (x_i)$.

Following the Bayesian methodology described in Ref.\cite{Neufcourt2020-limits}, we constructed emulators for separation energy residuals $\delta^{GP}(N,Z)$ using Gaussian Processes (GP)
$ \delta^{GP}(x) \sim \mathcal{GP} (\mu, k_{\eta,\rho} (x,x'))$
over the bi-dimensional domain $x$. The GP is characterized by its mean function and covariance kernel, taken respectively as a constant $\mu$ and squared-exponential covariance kernel $k_{\eta,\rho} (x, x') := \eta^{2}
e^{-\frac{(Z-Z')^2}{\rho_{Z}}-\frac{(N-N')^2}{\rho_{N}}}$, where $\rho_{Z}$ and
$\rho_{N}$ are the correlation ranges along the proton and neutron direction, respectively.
We add to the model a  term accounting for statistical uncertainties, 
assumed independent, identically distributed and scaled by a parameter $\sigma$. 
This yields
\begin{equation}\label{stat-model}
    y_i= \delta^{GP}(x_i) + \sigma \epsilon_i.
\end{equation}
Thus our GP model is parametrized by the 5-dimensional vector $\theta:=(\mu, \eta, \rho_Z, \rho_N, \sigma)$.

Posterior distributions for the $\mathcal{GP}$ parameters are obtained via Bayes' equation
\begin{equation}
	p(y|\theta):= \frac{p(\theta|y) \pi(\theta)}{\int p(\theta|y)
	                    \pi(\theta)d\theta}\,,
\end{equation}
where $p(\theta|y)$ is the statistical model \eqref{stat-model}'s likelihood and $\pi(\theta)$ the prior on its parameters. Priors we taken weakly informative, as described in Ref. \cite{Neufcourt2020-limits}. Samples from the posterior distributions of the $\mathcal{GP}$ parameters were drawn from iterations of a Monte Carlo Markov Chain. These samples of the residuals' emulators were in turn used to produce samples of two-proton separation energies and mass filters, as well as derive statistical predictions: averages and corresponding correlated uncertainties along with full covariance matrices. 

In a second stage of the analysis, we ensemble the emulators built from each individual nuclear model according to their BMA weights, namely the posterior probability for each model to be the hypothetical \emph{true} model, \emph{assuming it is one of them}, given priors on model weights and data. While the classical BMA literature\citemeth{KASS1995} relies on the same data $y$ as used for the individual model's training, for this step we prefer to use new ``testing'' data $y^*$ ($^{48}$Ni, $^{54}$Zn, $^{84}$Zr, $^{86}$Mo, $^{90}$Ru, $^{92}$Ru, and $^{94}$Pd) located at the outer boundary of the training set and excluded from the GP training. This ensures that the weights reflect better the extrapolative power of the models, and reduces overfitting. Formally we can write\cite{Neufcourt2020-limits} these BMA weights as 
\begin{equation}
	w_k = p(\mathcal{M}_k | y^*) = \frac{p(y^* | \mathcal{M}_k)\pi(\mathcal{M}_k)}%
	{\sum_{\ell=1}^{11} p(y | \mathcal{M}_\ell) \pi(\mathcal{M}_\ell)}\,,
	\label{eq:BMA_weights}
\end{equation}
where $\pi(\mathcal{M}_k)$ are prior model weights, 
and $p(y
|\mathcal{M}_k)$ are the model evidences obtained by integrating the likelihood
equation over the parameter space.  For our GP emulators, this gives 
\begin{equation}
	p(y | \mathcal{M}_k) = \int p(y|\theta_k, \mathcal{M}_k) \pi(\theta_k,
	\mathcal{M}_k) d\theta_k \,.
\end{equation}
We assume uniform prior weights, which are from a statistical standpoint the unique non-informative prior distribution in this setup. In order to speed up computations and increase stability\citemeth{Kejzlar2019BMA}, the evidence integrals are calculated using the Laplace approximation\citemeth{KASS1995}, where it is assumed that the posterior is Gaussian with the same mean and standard deviation. The resulting model evidences are:
\begin{equation}
	p(y | \mathcal{M}_k) \approx \exp\bigg [-\sum_i\frac{(y_i^{exp} - y^{(k)}(x_i))^2}{2\sigma_{y_k}(x_i)^2}\bigg] \,,
\end{equation}
where $y^{(k)}$ are the individual model emulators' predictions, $\sigma_{y_k}(x)$ the corresponding uncertainties, and $i$ runs over the retained set of nuclei\citemeth{Kejzlar2019BMA}. 

The model weights (rounded to two decimal digits) are: $w_k$=0.01 (SkM*),  0.04 (SkP), 0.12 (SLy4), 0.16 (SV-min), 0.07 (UNEDF0), 0.11 (UNEDF1), 0.20 (UNEDF2), 0.05 (BCPM), 0.21 (D1M), 0.00 (FRDM), and 0.00 (HFB-24). The final BMA predictions and uncertainties are calculated as 
\begin{align*}
	y(x) &= \sum_k w_k y^{(k)}(x) \,; \\
	\sigma_y^2(x) &= \sum_k w_k (y^{(k)}(x) - y(x))^2 + \sum_k w_k \sigma_{y_k}^2(x) \,,
\end{align*}
This last equation conveniently splits the uncertainties into the uncertainty on the model choice and the uncertainty on the individual models' parameters, and highlights what would be lost if a single model were used.

\begin{figure}[!htb]
        \includegraphics[width=1.0\linewidth]{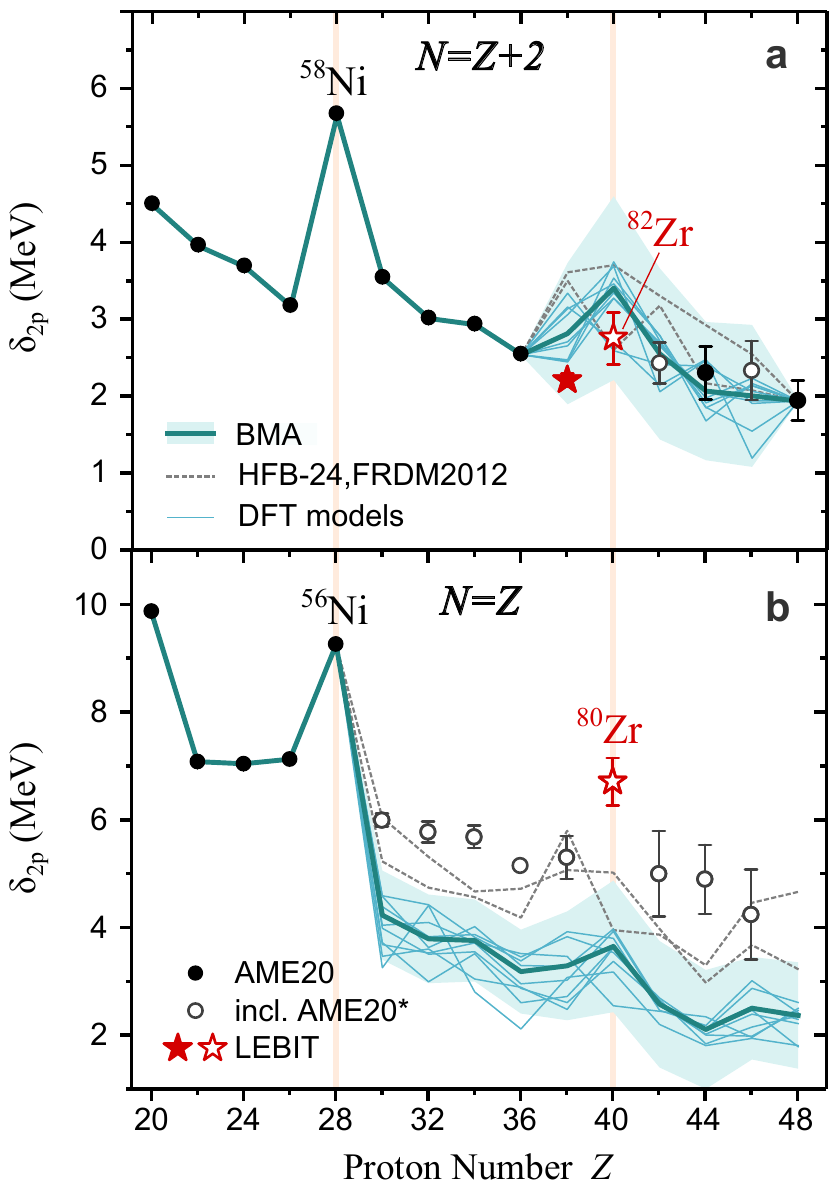}
        \caption{\textbf{Two-proton shell gap.} The effect of the anomalous mass of $^{80}$Zr on the mass indicator $\delta_{2p}$:  a  decrease from the baseline in the $N=Z+2$ sequence \textbf{(a)}, and a corresponding rise in the $N=Z$ sequence \textbf{(b)}. Black circles represent mass data from the AME20\cite{AME20}. Red stars include data from this work. Open symbols include mass extrapolations  from AME20\cite{AME20}.   
        The thick teal line is the BMA result based on several nuclear models (thin solid lines: DFT models; thin dashed lines: HFB-24 and FRDM2012 models that include the Wigner-energy correction), and the light teal band represents the uncertainty of the BMA approach. }
        \label{fig:graphsd2p}
\end{figure}

\textbf{\textcolor{blue}{The two-proton shell gap}}

Figure~\ref{fig:graphsd2p} displays the two-proton shell gap $\delta_{2p}$ (\ref {eq:Dpp}).  For the $N=Z+2$ sequence, the BMA prediction agrees with experiment within the estimated uncertainty. For $N=Z$, the anomalous mass of $^{80}$Zr results in an increase of  $\delta_{2p}$ above the baseline. Similar to what is seen in Figure 2b, the HFB-24 and FRDM2012 models that include the Wigner-energy correction lie slightly below the data points. As discussed earlier, this suggests that the Wigner energy term is underestimated by both models.

\bibliographystylemeth{naturemag}
\bibliographymeth{library}



\textbf{\textcolor{blue}{Data availability}}

The data that support the plots within this paper and other findings of this study are available from the corresponding author upon reasonable request.

\textbf{\textcolor{blue}{Code availability}}

Our unpublished computer codes  used to generate results that are reported in the paper and central to its main claims will be made  available upon request, to editors and reviewers.

\textbf{\textcolor{blue}{Acknowledgements}}

The authors would like to thank the NSCL staff for their technical support as well as Richard F. Casten for useful discussions on interpreting the results of the experiment. This work was conducted with the support of Michigan State University, the U.S. National Science Foundation under Contract No. PHY-1565546,  the U.S. Department of Energy, Office of Science, Office of Nuclear Physics under Award Nos. DE-SC0015927,
 DE-SC0013365, and  DE-SC0018083 (NUCLEI SciDAC-4 collaboration),  and by the National Science Foundation CSSI program under award number 2004601 (BAND collaboration).
 
\textbf{\textcolor{blue}{Author contributions}}

A.H., E.L., G.B., K.L., C.N., D.P., R.R., C.S.S., and I.T.Y. performed the experiment. A.H., E.L., D.P., and I.T.Y. performed the data analysis. A.H., E.L., W.N., S.A.G., and L.N. prepared the manuscript. R.J., S.A.G., W.N., and L.N. performed the Bayesian analysis. 
All authors discussed the results and provided comments on the manuscript.

\textbf{\textcolor{blue}{Competing interests}}

The authors declare that they have no competing financial interests.

\textbf{\textcolor{blue}{Correspondence and requests for materials}}

Correspondence and requests for materials should be addressed to A.H.~(email: hamaker@nscl.msu.edu).


\clearpage

\end{document}